\documentstyle[preprint,aps]{revtex}
\tightenlines
\begin{document}
\draft
\title
{Fractons and Fractal Statistics}
\author
{Wellington da Cruz\footnote{E-mail: wdacruz@exatas.uel.br}} 
\address
{Departamento de F\'{\i}sica,\\
 Universidade Estadual de Londrina, Caixa Postal 6001,\\  
Cep 86051-990 Londrina, PR, Brazil\\}
\date{\today}
\maketitle
\begin{abstract}

Fractons are anyons classified into equivalence 
classes and they obey a 
specific fractal statistics. The equivalence classes are labeled 
by a fractal parameter or Hausdorff dimension $h$. 
We consider this approach in the context of the Fractional Quantum 
Hall Effect ( FQHE ) and the concept of duality 
between such classes, defined by $\tilde{h}=3-h$ shows 
us that the filling factors for which the FQHE were 
observed just appear into these classes. A connection between 
equivalence classes $h$ and the modular group for the 
quantum phase transitions of the FQHE is also obtained. 
A $\beta-$function is defined for a complex conductivity 
which embodies the classes $h$. The thermodynamics is also 
considered for a gas of fractons $(h,\nu)$ with a constant 
density of states and an exact equation of state 
is obtained at low-temperature and low-density limits. 
We also prove that the Farey sequences for 
rational numbers can be expressed 
in terms of the equivalence classes $h$.

\end{abstract}



\section{Introduction}

We obtain from a continuous family of 
Lagrangians for fractional spin particles a path integral 
representation for the propagator and its representation in 
momentum space. In continuing, we consider the trajectories swept out 
by scalar and spinning particles which are characterized by 
the fractal parameter $h$ ( or the Hausdorff dimension ). 
We have that $L$, 
the length of closed trajectory with size $R$ has its fractal 
properties described by $L\sim R^h$, so 
for scalar particle $h=2$ and for  
 spinning particle $h=1$. From the form of 
 anyonic propagator obtained, with the spin defined 
 into the interval $0\leq s\leq \frac{1}{2}$, we have 
 extracted that the Hausdorff  dimension $h$ and the spin $s$ 
 of the particle are related by $h=2-2s$. Thus, 
 for anyonic particle, $h$ takes values within the interval 
 $1$$\;$$ < $$\;$$h$$\;$$ <$$\;$$ 2$ and the fractional 
spin particles or anyonic excitations are classified in terms of 
equivalence classes labeled by $h$. 
Therefore, such particles in a specific class 
can be considered on equal footing.

On the other hand, in the context of FQHE systems, the 
filling factor, 
a parameter which characterizes that phenomenon, 
can be classified in this 
terms too. We have, therefore, a new
hierarchy scheme for the filling factors, which is 
extracted from the relation between $h$ and the 
statistics $\nu$ 
( or the filling factors ). Our approach, contrary to 
literature does not have an empirical character 
and we 
can predict for which values of $\nu$ FQHE can be 
observed. 
A braid group structure behind this classification 
 is also obtained, such that the anyonic model, a 
charge-flux system, constitutes a topological obstruction 
in this description and the elements of the braid 
group are equivalent trajectories. 

Furthermore, we obtain a {\bf fractal statistics} ( the distributions 
have properties of {\it fractal functions} ) for fractional 
spin particles termed {\it fractons}$\;$$(h,\nu)$\footnote{Be aware, 
the word {\it fractons} was used before in the literature 
denoting distinct things, see references\cite{R33} 
and Plyushchay in\cite{R20}. In particular, the last 
one consider anyonic theory, however it has nothing to 
do with our definition.}  
, i.e. 
anyons classified into universal class $h$ of particles. As a result, 
{\it the Fermi-Dirac and Bose-Einstein statistics 
are generalized naturally}. A connection with FQHE is then done 
and the filling factors observed for such phenomenon appear 
into these equivalence classes $h$. The concept of duality between 
equivalence classes introduces the idea of supersymmetry 
in this context in a manner it looks like very interesting. 
This way, we have proved that the constraints that the 
filling factors obey for the quantum phase transitions 
between Hall plateaus takes place are satisfied 
by the classes $h$. Thus, we know all possible 
transitions since we have a fractal spectrum 
relating $h$ and $\nu$, i.e. $h$ classifies the 
universality class of the quantum 
Hall transitions. The fractal parameter $h$ works in 
this way as a critical exponent to the correlation lenght 
and our theoretical values coincide 
with the experimental ones.  As a consequence, a $\beta$-function 
for a complex conductivity is obtained in terms of $h$. 
The termodynamics for a free gas of fractons$\;$$(h,\nu)$ is 
obtained at low-temperature and low-density limits for 
a constant density of states.

Finally, a connection between number theory and 
Hausdorff dimension is established from our 
considerations about FQHE. The Farey sequences of rational 
numbers can be expressed in terms of  the fractal 
parameter $h$. We close 
this paper with a summary, enphasizing crucial 
theoretical points. All 
necessary background is posed along of our topic expositions.

\section{Anyonic Propagator}

There are various actions in the literature\cite{R1}
 for particles 
with fractional spins, called anyons, that can be 
present in nature and explain phenomena such as the
 fractional quantum
 Hall effect and high temperature 
superconductivity\cite{R2}. Such models have been
 constructed in at least 
two ways: one is coupling to some matter field a statistical 
 field of the 
Chern-Simons type that changes the statistics; and 
the other is the group-theoretical 
method that implies a classical mechanical model 
with a Poincar\'e invariant 
Lagrangian which is quantized\cite{R3,R4}. We 
consider a simple action 
obtained by the latter approach with minimal extension of 
the phase space that preserves all canonical structure of the 
space-time and spin algebra. 
Thus, we propose a path integral representation for the propagator 
in a  convenient gauge. In\cite{R3}, the 
following continuous family of 
Lagrangians for free anyons was obtained

\begin{eqnarray}
L_{s}=\frac{1}{2e}\left({\dot {x}_{\mu}}
-\frac{s\;{\dot q}\;
f^{\prime}_{\mu}}
{m\;f.\;{\dot x}}\right)^{2}+\frac{e\;m^2}{2}
-\frac{\alpha\; m\; e\;
 {\dot q}}{f.\;{\dot x}},
\end{eqnarray}

\noindent  where $ m $ and $\alpha$ are 
respectively the mass and
the helicity 
of the anyon, and $ s$ is an arbitrary constant
 deduced from the 
Casimir of the spin algebra, $ S^2=-s^2$.
 The functions $f_{\mu}$ satisfy the non-linear
  differential equation
 
 \begin{equation}
 \label{w1}
 f_{\mu}=\epsilon_{\mu\alpha\beta}f^{\prime\alpha}f^{\beta}
 \end{equation}
 
 \noindent and we obtain the relations
 
 \begin{equation}
 f^2=0,\;f^{\prime 2}=-1,\;f^{\prime\prime}.\;f=1,
 \end{equation}
 
 \noindent with $ f^{\prime}=\frac{df}{dq}$. On the
  other hand,
  we have that 
 $ S_{\mu}=\pi\;f_{\mu}+s\;f_{\mu}^{\prime}$, so we get,
  $ S_{\mu}\;
 f^{\prime\mu}=-s$,$\;$ $ S_{\mu}\;f^{\mu}=0$,$\;$
 $ S_{\mu}^{\prime}\;f^{\mu}=s$,$\;$
 $ S_{\mu}^{\prime}\;f^{\prime\mu}=-\pi$. The equation $
  {\dot f}-{\dot q}f^{\prime}=0$ is equivalent to eq.(\ref{w1})
   and 
 \[ 
 {\dot f}^{\prime}
 -\frac{\partial{\dot q}}{\partial{q}} f^{\prime}
 -{\dot q}\frac{\partial{f^{\prime}}}{\partial{q}}=0, 
 \] 
 and for $f_{\mu}=a_{\mu}q^2+b_{\mu}q+c_{\mu}$, we have 
 $a_{\mu}=(1,0,-1)
 $,$\;$$b_{\mu}=(-1,1,1)$,$\;$$c_{\mu}=\frac{1}{2}(1,-1,0)$.
  This way,
  we extract for $q$ not constant, the condition
   $\frac{\partial{\dot q}}
  {\partial{q}}=0$. The Lagrangian 
 is subject to the constraint, $ \sigma\;(f.\;{\dot x})=e\;{\dot q}$,
  and with 
 ${\dot x}_{\mu}=e\;p_{\mu}+\sigma \;S_{\mu}$, we obtain 
 $ {\dot q}=\sigma \; (f\;.\;p)$ 
 which satisfies and gives us that $ f^{\prime}\;.\;p=0$. Then, 
 $ {\dot x}\;.\;f^{\prime}=-\sigma \;s$ and 
 $ \pi \;(f\;.\;p)=-\alpha \; m$, 
 i.e. $ S_{\mu}\;p^{\mu}+\alpha m=0$ and $ {\dot x}\;.\;f
 =e\;f^{\mu}\;p_{\mu}$, and in the gauge
  ${\dot {\sigma}}=0,\;\
 {\dot {e}}=0$, we have ${\dot q}=m\;f.\;{\dot x}$,
  so $ \sigma=e\;m$
  and  $ p^2-m^2=0 $.  
 
 Now, we consider the path integral 
 representation for 
 the anyonic propagator with the action, ${\cal A}=
 \int_{0}^{1} L_{s}\;d\tau$.
 In that gauge, we verify 
 that the continuous family of Lagrangians reads
  
  \begin{equation}
  \label{a14}
  L_{s}=\frac{{\dot x}^2}{2e}+\frac{s^2}{2e}
  +\frac{e}{2}(m^2-2m^2\alpha).
  \end{equation}
  
  This way, we propose the path integral 
  representation for the free fractional spin particle as
 
\begin{eqnarray}
\label{p1}
{\cal F}(x_{out},x_{in})=&&\frac{\imath}{2}\int_{0}^{\infty} 
de_{0}\int_{x_{in}}^{x_{out}}
{\cal D}x\;{\cal D}e\; M(e)\;\delta({\dot e})\\
&&\times\exp\left\{-\frac{i}{2}\int_{0}^{1}
\left[\frac{{\dot x}^2}{e}+
e{\cal M}^2+\frac{s^2}{e}\right]
d\tau\right\},\nonumber
\end{eqnarray}

\noindent where $ {\cal M}^2=m^2-2m^2\alpha$ and the delta
 functional takes into account the gauge ${\dot{e}}=0$. 
We define the measure

\begin{eqnarray}
M(e)=\int {\cal D}p\;\exp\left\{\frac{\imath}{2}
\int_{0}^{1}e\;p^2\; 
d\tau\right\},
\end{eqnarray}

\noindent such that, making the substitution

\[
{\sqrt {e}}p\rightarrow p, \; 
\frac{x-x_{in}-\tau{\Delta x}}{\sqrt e} \rightarrow x,\;
{\Delta x}=x_{out}-x_{in},\nonumber
\]
\noindent we obtain the new boundary conditions
 $x(0)=0=x(1)$. So, 
in eq.(\ref{p1}), we have the factor

\begin{eqnarray}
I=&&\frac{\imath}{2}\int_{0}^{0} {\cal D}x\; {\cal D}p\; 
\exp\left\{\frac{\imath}
{2}\int_{0}^{1}(p^2-{\dot x}^2)d\tau\right\}\nonumber\\
=&&\sqrt{\frac{\imath}{4(2\pi)^3}}.
\end{eqnarray}

\noindent Given that

\[
\int {\cal D}e\; \delta\left({\dot e}\right){\cal G}(e)
={\cal G}(e_{0}),
\]

\noindent we arrive at the path integral representation for 
the propagator
\begin{eqnarray}
\label{a12}
&& {\cal F}(x_{out},x_{in})=\sqrt{\frac{\imath}{4(2\pi)^3}}
\int_{0}^{\infty}
\frac{de_{0}}{\sqrt{e_{0}^3}}\exp\left\{-\frac{\imath}{2}
\left[e_{0}
{\cal M}^2+\frac{{\Delta x}^2}{e_{0}}-\frac{s^2}
{e_{0}}\right]
\right\}
\end{eqnarray}

\noindent and $s^2>0$ for the causal propagator. This 
expression 
was obtained along the same lines as those used elsewhere 
for the scalar particle\cite{R5}. On 
the other hand, in momentum space, the anyonic propagator 
has the form ( da Cruz in\cite{R6} )

\begin{equation}
\label{a7}
{\tilde{\cal F}}(p)=\frac{1}{(p_{\mu}-mS_{\mu})^2-m^2}.
\end{equation}

\section{Hausdorff dimension and anyonic excitations}

The trajectories swept out 
by scalar and spinning particles can be characterized by 
the fractal parameter $h$ ( or the Hausdorff dimension ). 
We have that $L$, the length of closed trajectory with 
size $R$ has its fractal 
properties described by $L\sim R^h$ ( see Polyakov; Kr\"oger 
in\cite{R6} ), so 
for scalar particle $L\sim \frac{1}{p^2}$, 
$R^2\sim L$ and $h=2$ ; for  
 spinning particle, $
 L\sim \frac{1}{p}$, $R^1\sim L$ and $h=1$. Thus, 
 for anyonic particle, $h$ takes values within the interval 
 $1$$\;$$ < $$\;$$h$$\;$$ <$$\;$$ 2$ and the fractional 
spin particles or anyonic excitations are classified in terms of 
equivalence classes labeled by $h$. From the form of 
 anyonic propagator given in eq.(\ref{a7}), with the spin defined 
 into the interval $0\leq s\leq \frac{1}{2}$, we extract the 
 following formula $h=2-2s$, which relates the Hausdorff 
 dimension $h$ and the spin $s$ of the particle. This formula 
 can be generalized and considering {\it ab initio} 
 the spin-statistics relation $\nu=2s$, we obtain the 
 {\it fractal spectrum} 
 
\begin{eqnarray}
\label{e34}
&&h-1=1-\nu,\;\;\;\; 0 < \nu < 1;\;\;\;\;\;\;\;\;
 h-1=\nu-1,\;
\;\;\;\;\;\; 1 <\nu < 2;\;\nonumber\\
&&h-1=3-\nu,\;\;\;\; 2 < \nu < 3;\;\;\;\;\;\;\;\;
h-1=\nu-3,\;\;\;\;\;\;\; 3 < \nu < 4;\;\nonumber\\
&&h-1=5-\nu,\;\;\;\; 4 < \nu < 5;\;\;\;\;\;\;\;\;
h-1=\nu-5,\;\;\;\;\;\;\; 5 < \nu < 6;\;\\
&&h-1=7-\nu,\;\;\;\; 6 < \nu < 7;\;\;\;\;\;\;\;\;
h-1=\nu-7,\;\;\;\;\;\;\; 7 < \nu < 8;\;\nonumber\\
&&h-1=9-\nu,\;\;\;\;8 < \nu < 9;\;\;\;\;\;\;\;\;
h-1=\nu-9,\;\;\;\;\;\;\; 9 < \nu < 10;\nonumber\\
&&etc,\nonumber
\end{eqnarray}

\noindent and so, we have a mirror symmetry 
for all spectrum of $\nu$.

\section{ Hausdorff dimension and filling factors }

We can establish a connection between the 
fractal parameter $h$ and the {\it filling factor} $\nu$,
a parameter that appears in the context of the FQHE  
and for them we have experimental values\cite{R7}. This 
connection can be done once the anyonic model has 
been considered to explain that phenomenon\cite{R8}.

The FQHE is associated with a planar charged system in a 
perpendicular magnetic field and as a result a new type of correlated 
ground state occurs. The Hall resistance develops 
plateaus at quantized values in the vicinity of the 
filling factor or statistics $\nu$, which is related 
to the fraction of electrons that forms collective 
excitations as quasiholes or quasiparticles. Excitations 
above the Laughlin ground state are characterized, 
therefore, by $\nu$ and so we propose a 
{\it new hierarchy scheme} for the FQHE which gives us the 
possibility of {\it predicting} for which values 
of $\nu$ FQHE can be observed ( see below, section VI.A, the 
concept of duality between  equivalence classes ).
 
Our scheme is based on the intervals of definition 
of spin $s$, for fractional spin particles which are related 
to the Hausdorff dimension $h$. We verify that for some 
experimental values of $\nu$ for which the FQHE 
were observed the Hausdorff dimension is a rational 
number with an odd denominator 
( like the filling factor ). Thus, bearing in 
mind the condition, $1< h <2$, we can determine 
for which values of $h$, the statistics takes 
a specific value in eq.(\ref{e34}). For example, 
this {\it hierarchy scheme} gives us  for some 
experimental values of $\nu$ the respective 
values of $h$, 

\begin{eqnarray}
&&( h\;,\; 0< \nu < 1):\\
&&\left(\frac{9}{5},\frac{1}{5}\right),\;\left(\frac{12}{7},
\frac{2}{7}\right),\; 
\left(\frac{5}{3},\frac{1}{3}\right),\;
\left(\frac{8}{5},\frac{2}{5}\right),\nonumber\\
&& \left(\frac{11}{7},\frac{3}{7}\right),\;
\left(\frac{14}{9},\frac{4}{9}\right),\;
\left( \frac{13}{9},\frac{5}{9}\right),\;
\left( \frac{10}{7},\frac{4}{7}\right),\nonumber\\
&&\left(\frac{7}{5},\frac {3}{5}\right),\;
\left(\frac{4}{3},\frac{2}{3}\right),\; 
\left(\frac{6}{5},\frac{4}{5}\right);\nonumber\\ 
&&( h\;,\; 1 < \nu < 2 ):\\
&&\left(\frac{5}{3},\frac{5}{3}\right),\;
\left(\frac{13}{7},\frac{13}{7}\right),\;
\left(\frac{13}{9},\frac{13}{9}\right),\nonumber\\
&&\left(\frac{10}{7},\frac{10}{7}\right),\;
\left(\frac{7}{5},\frac{7}{5}\right),\;
\left(\frac{4}{3},\frac{4}{3}\right);\nonumber\\
&&( h\;,\; 2 < \nu < 3 ):\\
&&\left(\frac{4}{3},\frac{8}{3}\right),\;
\left(\frac{5}{3},\frac{7}{3}\right).\nonumber
\end{eqnarray}

\noindent We can see another 
interesting point from these pairs of numbers: Some 
collective excitations with different spins have the same 
value of $h$, i.e. the nature of the occurrence 
of FQHE for that values of $\nu$ can be classified in terms 
of $h$, so we can say that this number 
classifies {\it the collective excitations in 
terms of its homotopy class}\cite{R8}.    

\section{Filling factors and braid group}

As we pointed out the {\it Hausdorff dimension} 
$h$, can be considered as a parameter which classifies 
the {\it equivalence classes} 
of the {\it collective excitations} which occurs in the context 
of the FQHE. The elements of each equivalence class are the {\it 
filling factors or statistics} $\nu$, which characterize the 
collective excitations manifested as quasiholes or 
quasiparticles. Now, we propose rules of composition 
for these elements and we extract a {\it braid group} 
structure from this new hierarchy scheme for the filling factors. 

We know that for path integration on multiply connected 
spaces\cite{R9}, we need to assign different weights $\chi$, to 
homotopically disconnected paths. So, in terms of our scheme, 
as we just said, $h_{i}$ ( $i$ means a specific interval ), 
labels these classes, which form a group, 
the fundamental group $\Pi_{1}({\cal P})$, with elements 
$\chi=\exp
\left\{\imath \nu \varphi\right\}$. We also have that, for the 
abelian representation of the group, 
the weights satisfy the constraints\cite{R9}

\begin{eqnarray}
&&|\chi\left\{\nu_{i}\right\}_{h_{i}}|=1,\\
&&\chi\left\{\nu_{i}\right\}_{h_{i}}\chi\left\{\nu_{j}
\right\}_{h_{j}}=
\chi\left\{\nu_{i}\circ\nu_{j}\right\}_{h_{i\circ j}}.\nonumber
\end{eqnarray}

\noindent The weights $\chi$ represent a phase which 
we assign to the paths that contribute to the propagator
 
\begin{eqnarray}
{\cal K}(q^{\prime},t^{\prime};q,t)=
\sum_{\left\{\nu\right\}_{h}}\chi
\left\{\nu\right\}_{h}{\cal K}^{\left\{\nu\right\}_{h}}
(q^{\prime},t^{\prime};q,t).
\end{eqnarray}
 
\noindent We observe that, for the composition law, 
for example, with $\nu_{1}+\nu_{2}$, we do not consider 
integers since $\nu$ can only be a rational number with 
odd denominator ( see again the intervals of $\nu$ ). On the 
other hand, we verify that some compositions with elements 
of the same class give another element either in or out 
of the class and the same occurs for the distinct classes. 
But this does not matter, because we only consider paths 
within the same class for the path integration. For example, 
consider the classes
   
\begin{eqnarray}
\left\{\frac{2}{3},\frac{8}{3},\cdots
\right\}_{h=\frac{4}{3}};\;\;
\left\{\frac{3}{5},\frac{7}{5},\cdots
\right\}_{h=\frac{7}{5}}.
\end{eqnarray}

\noindent If we compose, $\left\{\frac{2}{3}
\right\}_{h=\frac{4}{3}}$ with 
$\left\{\frac{3}{5}\right\}_{h=\frac{7}{5}}$
 we get $\left\{
\frac{19}{15}\right\}_{h=\frac{19}{15}}$; if
 we compose $\left
\{\frac{2}{3}\right\}_{h=\frac{4}{3}}$ with
 $\left\{\frac{8}{3}
\right\}_{h=\frac{4}{3}}$ we get $\left\{\frac{10}{3}
\right\}
_{h=\frac{4}{3}}$; if we compose $\left\{\frac{3}{5}
\right\}_{h=\frac{7}{5}}$ with $\left\{\frac{7}{5}
\right\}_{h=\frac{7}{5}}$ we get $\left\{ 2\right\}$
 which is not defined, 
etc.

\section{Fractal statistics}

In the literature, the fractional spin particles are a subject 
of attention by now for a long date. Anyons\cite{R10} are 
kinds of particles, that live in two dimensions, where 
wavefunctions acquire an arbitrary phase when two of them are braided. 
A fractional exclusion statistics was proposed in\cite{R11} 
for arbitrary dimensions, which interpolates between bosonic 
and fermionic statistics. In\cite{R12} was showed that anyons 
also satisfy an exclusion principle when Haldane statistics are 
generalised to infinite dimensional Hilbert spaces. The statistics 
parameter $g$ was introduced  for particles with finite dimensional 
Hilbert spaces and defined as $g=-\frac{\Delta{D}}{\Delta{N}}$, 
and if 
we add $\Delta{N}$ particles into the system, the number 
of single-particles states $D$ available for further 
particles is altered by $\Delta{D}$. This expresses a 
reduction of single-particle Hilbert space by the exclusion 
statistics parameter $g$. In\cite{R13} a distribution function for 
particles obeying a generalised exclusion principle was introduced, 
and the statistical mechanics and thermodynamics considered. 
Also in\cite{R14}, some physical properties considering 
exclusion statistics for identical particles 
called $g$-ons was obtained.

On the other hand, in the context of Fractional 
Quantum Hall Effect wavefunctions 
were obtained for the lowest 
Landau level ( LLL)\cite{R15} and FQHE quasiparticles 
as anyons confined in the LLL were discussed in terms of exclusion 
statistics in\cite{R16}.

In this paper, {\it we propose another way to consider anyons}. 
As we have pointed out above they can be classified 
in terms of a fractal parameter or Hausdorff 
dimension $h$, and so these particles appear into 
the equivalence classes labeled by $h$. Contrary to 
literature\cite{R13} we show that different particles 
with distinct values 
of fractional spin $s$ within the equivalence class $h$ 
satisfy a specific fractal statistics. This constitutes
 in some sense a more suitable approach to 
 fractional spin particles, since unify them naturally. 
 A connection with FQHE systems can  be done, 
 and the filling factors for which FQHE were 
observed just appear into these classes and the ocurrence 
of FQHE are 
considered in terms of duality between equivalence classes. A 
supersymmetric character between fermions and bosons 
( dual objects by definition ) in the extremes 
of the range of $h$ is a special result 
advanced by this formulation.

The manifest symmetry of the relation 
between $h$ and $\nu$ ( see eq.(\ref{e34}) ) and the ways 
of distributing $N_{j}$ 
identical particles among $G_{j}$ states, besides the 
constraints given by the extremes of 
the interval of definition of $h$, i.e. $h=2$ for 
bosons and $h=1$ for fermions or in another way how 
distributing fermions and bosons bring us to 
propose a statistical weight for the anyonic excitations in terms 
of $h$, as follows:
  
\begin{equation}
\label{e.15}
\omega_{j}=\frac{\left[G_{j}+(N_{j}-1)(h-1)\right]!}{N_{j}!
\left[G_{j}+(N_{j}-1)(h-1)-N_{j}\right]!}.
\end{equation}

\noindent For $1$$\;$$ < $$\;$$h$$\;$$ <$$\;$$ 2$, 
we have anyonic 
excitations which interpolates between these two extremes. 
Consider now the classes $h=\frac{5}{3}$ and $h=\frac{4}{3}$, 
then we obtain
 $\left\{\frac{1}{3},\frac{5}{3},\frac{7}{3},
 \frac{11}{3},\frac{13}{3},\frac{17}{3},\cdots\right\}_
{h=\frac{5}{3}}$ and  $\left\{\frac{2}{3},
\frac{4}{3},\frac{8}{3},\frac{10}{3},\frac{14}{3},
\frac{16}{3},
\cdots\right\}_
{h=\frac{4}{3}}$.

These particles or excitations, in the class, share some 
common 
characteristics, according to our interpretation, in the 
same way that bosons and fermions. We observe again that 
each interval contributes with one and only one particle to the 
class and the statistics and the spin, are related by $\nu=2s$. 

We stress that, our expression for statistical 
weight 
eq.(\ref{e.15}), is more general than an expression given 
in\cite{R13},
once if we consider the relation between $h$ and $\nu$, we 
just obtain that first result and extend it for the 
complete spectrum of statistics $\nu$, for example: 
 
\begin{eqnarray}
\label{e.4}
&&\omega_{j}=\frac{\left[G_{j}+(N_{j}-1)(1-\nu)\right]!}
{N_{j}!\left[G_{j}+(N_{j}-1)(1-\nu)-N_{j}\right]!},
 \;\;\;\; 0 < \nu < 1; \nonumber\\
&&\omega_{j}=\frac{\left[G_{j}+(N_{j}-1)(\nu-1)\right]!}
{N_{j}!\left[G_{j}+(N_{j}-1)(\nu-1)-N_{j}\right]!},
\;\;\;\; 1 < \nu < 2; \\
&&\omega_{j}=\frac{\left[G_{j}+(N_{j}-1)(3-\nu)\right]!}
{N_{j}!\left[G_{j}+(N_{j}-1)(3-\nu)-N_{j}\right]!},
\;\;\;\; 2 < \nu < 3;\nonumber\\
&&\omega_{j}=\frac{\left[G_{j}+(N_{j}-1)(\nu-3)\right]!}
{N_{j}!\left[G_{j}+(N_{j}-1)(\nu-3)-N_{j}\right]!},
\;\;\;\; 3 < \nu < 4;\nonumber\\
&&etc.\nonumber
\end{eqnarray}

\noindent  The expressions 
eq.(\ref{e.4}) were possible because of the mirror symmetry 
in eq.(\ref{e34}). But, our approach 
in terms of Hausdorff dimension $h$, have an advantage, 
because we collect into equivalence class the anyonic 
excitations and so, we consider on equal footing the 
excitations 
in the class. We have, therefore, a new approach 
for fractional spin particles. For different 
species of particles, the statistical weight take the form

\begin{equation}
\label{e.5}
\Gamma=\prod_{j}\omega_{j}=\prod_{j}\frac{\left
[G_{j}+(N_{j}-1)
(h-1)\right]!}{N_{j}!\left[G_{j}+(N_{j}-1)(h-1)-N_{j}\right]!},
\end{equation}
\noindent which reduces to eq.(\ref{e.15}) for 
only one species. Now, we can consider the entropy 
for a given class $h$. Taking the logarithm of 
eq.(\ref{e.5}), 
with the condition that, $N_{j}$ and $G_{j}$ are vary 
large numbers and 
defining the average occupation numbers, $n_{j}=
\frac{N_{j}}{G_{j}}$, 
we obtain for a gas not in equilibrium, an expression 
for the entropy

\begin{eqnarray}
&&{\cal S}=\sum_{j}G_{j}\left\{\left[1+n_{j}(h-1)
\right]\ln\left(
\frac{1+n_{j}(h-1)}{1+n_{j}(h-1)-n_{j}}\right)\right.\\
&&-\left.n_{j}\ln\left(\frac{n_{j}}{1+n_{j}(h-1)-n_{j}}
\right)\right\},
\nonumber
\end{eqnarray}

\noindent and for $h=2$ and $h=1$, we obtain
the expressions for a Bose and Fermi 
gases not in equilibrium\cite{R17}, 
  
\begin{eqnarray}
&&{\cal S}=\sum_{j}G_{j}\left\{\left(1+n_{j}\right)\ln
\left(1+n_{j}\right)-n_{j}\ln n_{j}\right\};\\
&&{\cal S}=-\sum_{j}G_{j}\left\{n_{j}\ln n_{j}+
\left(1-n_{j}\right)\ln \left(1-n_{j}\right)\right\},
\end{eqnarray}
\noindent respectively.

The distribution function for a gas of the class $h$ can 
be obtained from the condition of the entropy be a 
maximum. Thus, 
we have

\begin{eqnarray}
\label{e.9}
n_{j}\xi=\left\{1+n_{j}(h-1)\right\}^{h-1}\left\{1+n_{j}
(h-2)\right\}^{2-h}
\end{eqnarray}

\noindent or 
\begin{eqnarray}
\label{e.45} 
n_{j}=\frac{1}{{\cal{Y}}[\xi]-h},
\end{eqnarray}

\noindent where the function ${\cal{Y}}[\xi]$ satisfies 
\begin{eqnarray}
\label{e.46} 
\xi=\left\{{\cal{Y}}[\xi]-1\right\}^{h-1}
\left\{{\cal{Y}}[\xi]-2\right\}^{2-h},
\end{eqnarray}
 
 \noindent and $\xi=\exp\left\{(\epsilon_{j}-\mu)/KT\right\}$, 
has the usual definition. The Bose and Fermi distributions
 are obtained for values of $h=2,1$ respectively. 
 At this point, 
 we observe that the condition of periodicity on 
 the statistics $\nu$, in our 
 approach, is expressed as 
 
 \begin{equation}
 n_{j}(\nu)=n_{j}(\nu+2),
 \end{equation}
 
 \noindent and the equivalence classes $h$ always respect 
 this condition naturally. 
 On the other hand, this also means that, as $\nu=2s$, particles 
 with distinct values of spin $s$ into the class $h$ obey a 
 specific fractal statistics eq.(\ref{e.45}).\footnote{
 We observe here that these distributions just as defined are {\it 
 fractal functions} which satisfy the necessary conditions as 
 discussed, 
 for example, by Rocco and West in\cite{R6}.} If the temperature 
 is sufficiently low and $\epsilon_{j}<\mu$, we can check 
 that the mean ocuppation number is given 
 by $n=\frac{1}{2-h}$, and so the fractal parameter $h$ 
 regulates the number of particles in each quantum state, i.e. 
 for $h=1$,$\;$$n=1$;$\;$$h=2$,$\;$$n=\infty$;$\;$$h=
 \frac{3}{2}$,$\;$$n=2$; etc. At $T=0$ and $\epsilon_{j}>\mu$, $n=0$
 if $\epsilon_{j}>\epsilon_{F}$ and $n=\frac{1}{2-h}$ 
 if $\epsilon_{j}<\epsilon_{F}$, hence we get a step distribution, 
 taking into account the Fermi energy $\epsilon_{F}$ and $h\neq 2$.
 
 \subsection{Fractional quantum Hall effect}

In the 
context of the 
Fractional Quantum Hall Effect, as we have pointed out above, 
the filling factor ( rational number with an 
odd denominator ), 
can be also classified in 
terms of $h$. We have that the anyonic excitations 
are collective 
excitations manifested as quasiparticles or quasiholes in FQHE 
systems. 
Thus, for example, we have the collective 
excitations as given in 
the beginning for $h=\frac{5}{3}$ and 
$h=\frac{4}{3}$. Now, we 
have noted 
that these collections contain filling fractions, in 
particular, $\nu=
\frac{1}{3}$ and $\nu=\frac{2}{3}$, that experimentally 
were observed\cite{R7} and so we are 
able to estimate for which values of $\nu$ the largest 
charge gaps occurs, or alternatively, we can {\it predict} 
FQHE. As was 
observed before this is a {\bf new hierarchy scheme} for 
the filling 
factors, that {\bf expresses the occurrence of the} FQHE {\bf in more 
fundamental terms}, i.e. {\bf relating the fractal parameter} $h$ 
{\bf and the filling factors} $\nu$\footnote{The quantum number $\nu$ gets its 
topological character from $h$.}. Of course, in our approach 
we do not have empirical expressions like this one\cite{R18}, 
$\nu=\frac{n}{2pn\pm 1}$, for the experimental occurrence 
of FQHE. After all, for 
anyonic excitations in a stronger 
magnetic field 
at low temperature, our results 
eq.(\ref{e.4}) obtained via an approach completely 
distinct of\cite{R13} can be 
considered. 

In another way, we introduce the concept of duality between 
equivalence classes, defined by 

\begin{equation} 
\label{e.10}
\tilde{h}=3-h,
\end{equation}

\noindent such that for $h=1$, $\tilde{h}=2$ and 
for $h=2$, $\tilde{h}=1$, so fermions and bosons 
are dual objects. This means that they can be 
considered supersymmetric particles\footnote{There are in the literature 
different formulations relating 
duality and FQHE\cite{R19}.}. 

For a set of values of the filling factors 
$\nu$ experimentally observed\cite{R7}, according 
to our relations (eqs.(\ref{e.10}) and (\ref{e34})), 
we get the 
classes, $h$ and $\tilde{h}$:

\begin{eqnarray}
&&\left\{\frac{1}{3},\frac{5}{3},\frac{7}{3},
 \frac{11}{3},\cdots\right\}_
{h=\frac{5}{3}},\;\;\;\;\;\;\;\;\;\;\left\{\frac{2}{3},
\frac{4}{3},\frac{8}{3},\frac{10}{3},\cdots\right\}_
{{\tilde{h}}=\frac{4}{3}};\nonumber\\
&&\left\{\frac{1}{5},\frac{9}{5},\frac{11}{5},
 \frac{19}{5},\cdots\right\}_
{h=\frac{9}{5}},\;\;\;\;\;\;\;\;\left\{\frac{4}{5},
\frac{6}{5},\frac{14}{5},\frac{16}{5},\cdots\right\}_
{{\tilde{h}}=\frac{6}{5}};\nonumber\\
&&\left\{\frac{2}{7},\frac{12}{7},\frac{16}{7},
 \frac{26}{7},\cdots\right\}_
{h=\frac{12}{7}},\;\;\;\;\;\left\{\frac{5}{7},
\frac{9}{7},\frac{19}{7},\frac{23}{7},\cdots\right\}_
{{\tilde{h}}=\frac{9}{7}};\nonumber\\
&&\left\{\frac{2}{9},\frac{16}{9},\frac{20}{9},
 \frac{34}{9},\cdots\right\}_
{h=\frac{16}{9}},\;\;\;\;\;\left\{\frac{7}{9},
\frac{11}{9},\frac{25}{9},\frac{29}{9},\cdots\right\}_
{{\tilde{h}}=\frac{11}{9}};\nonumber\\
&&\left\{\frac{2}{5},\frac{8}{5},\frac{12}{5},
 \frac{18}{5},\cdots\right\}_
{h=\frac{8}{5}},\;\;\;\;\;\;\;\;\left\{\frac{3}{5},
\frac{7}{5},\frac{13}{5},\frac{17}{5},\cdots\right\}_
{{\tilde{h}}=\frac{7}{5}};\\
&&\left\{\frac{3}{7},\frac{11}{7},\frac{17}{7},
 \frac{25}{7},\cdots\right\}_
{h=\frac{11}{7}},\;\;\;\;\;\left\{\frac{4}{7},
\frac{10}{7},\frac{18}{7},\frac{24}{7},\cdots\right\}_
{{\tilde{h}}=\frac{10}{7}};\nonumber\\
&&\left\{\frac{4}{9},\frac{14}{9},\frac{22}{9},
 \frac{32}{9},\cdots\right\}_
{h=\frac{14}{9}},\;\;\;\;\;\left\{\frac{5}{9},
\frac{13}{9},\frac{23}{9},\frac{31}{9},\cdots\right\}_
{{\tilde{h}}=\frac{13}{9}};\nonumber\\
&&\left\{\frac{6}{13},\frac{20}{13},\frac{32}{13},
 \frac{46}{13},\cdots\right\}_
{h=\frac{20}{13}},\;\;\;\left\{\frac{7}{13},
\frac{19}{13},\frac{33}{13},\frac{45}{13},\cdots\right\}_
{{\tilde{h}}=\frac{19}{13}};\nonumber\\
&&\left\{\frac{5}{11},\frac{17}{11},\frac{27}{11},
 \frac{39}{11},\cdots\right\}_
{h=\frac{17}{11}},\;\;\;\left\{\frac{6}{11},
\frac{16}{11},\frac{28}{11},\frac{38}{11},\cdots\right\}_
{{\tilde{h}}=\frac{16}{11}};\nonumber\\
&&\left\{\frac{7}{15},\frac{23}{15},\frac{37}{15},
 \frac{53}{15},\cdots\right\}_
{h=\frac{23}{15}},\;\;\;\left\{\frac{8}{15},
\frac{22}{15},\frac{38}{15},\frac{52}{15},\cdots\right\}_
{{\tilde{h}}=\frac{22}{15}}.\nonumber
\end{eqnarray}

\noindent We {\it emphasize} that in each class, some 
filling factors are just the experimental values 
observed, i.e. the Hall resistance  develops plateaus in these 
quantized values, which are related to the fraction of 
electrons that form collective excitations as 
quasiholes or quasiparticles in FQHE systems. The 
relation 
of duality between equivalence classes labeled by $h$ 
can, therefore, indicates a way as determine the dual 
of a specific value of $\nu$ ( or ${\tilde{\nu}}$ ) 
observed. Note also that the class $
\left\{1,3,5,7,\cdots\right\}_{h=1}$ 
gives us the odd filling factors 
for the integer quantum Hall effect. An interpretation 
of the fractal parameter $h$ as some kind of {\it 
order parameter} which characterizes the ocurrence of FQHE 
is enough attractive. We can say that the 
concept of {\it duality} connects a {\it 
quasi-bosonic regime} $h\sim 2$ to a {\it 
quasi-fermionic regime} $h\sim 1$, as $(h,\nu)=
(\frac{5}{3},\frac{1}{3})$ to $(\frac{4}{3},\frac{2}{3})$. 
This observation overlapping with some ideas in the 
literature for considering supersymmetric anyon theories in 
condensed matter systems\cite{R20}. As a check we can verify 
that the supersymmetric pairs for energy states of spin 
$\left(s,s\pm\frac{1}{2}\right)$ pointed out by 
Spector, and Witten\footnote{ Here, in particular, we are 
suggesting a connection with {\it fractal 
superstrings}, see Tye in\cite{R20}.}\cite{R20} have a  realization  
through duality concept introduced here\footnote{For another discussion about 
deformed Lie algebras and anyonic supersymmetry, where systems 
of two states have spins shifted in one-half, see Plyushchay in \cite{R20}.}
. Therefore, the 
FQHE is a natural scenario for {\it fractal supersymmetry} 
to be explored ( see Traubenberg and Spulinski, and more in\cite{R20} ).

Another view is the relation $L\sim R^h$, which shows us a 
{\it scaling law} behind this characterization of the FQHE\cite{R6}. 
We also observe that our approach, in terms of equivalence 
classes for the filling factors, embodies 
the structure of the {\it modular group} as discussed 
in\cite{R19,R21}. For 
that, we consider the sequences given by Dolan\cite{R21} as

\begin{eqnarray}
&&(h,\nu)=\left(\frac{5}{3},\frac{1}{3}\right)\rightarrow 
\left(\frac{8}{5},
\frac{2}{5}\right)
\rightarrow \left(\frac{11}{7},\frac{3}{7}\right)\rightarrow 
\left(\frac{14}{9},\frac{4}{9}\right)
\rightarrow \left(\frac{17}{11},\frac{5}{11}\right)
\rightarrow 
\left(\frac{20}{13},\frac{6}{13}\right) \rightarrow 
\cdots;\nonumber\\
&&(h,\nu)=\left(\frac{19}{13},\frac{7}{13}\right)\rightarrow 
\left(\frac{16}{11},\frac{6}{11}\right)
\rightarrow \left(\frac{13}{9},\frac{5}{9}\right)
\rightarrow \left(\frac{10}{7},\frac{4}{7}\right)
\rightarrow \left(\frac{7}{5},\frac{3}{5}\right) \rightarrow 
\left(\frac{4}{3},\frac{2}{3}\right)\rightarrow (1,1);\\
&&(h,\nu)=\left(\frac{4}{3},\frac{2}{3}\right)\rightarrow 
\left(\frac{9}{7},\frac{5}{7}\right)
\rightarrow \left(\frac{14}{11},\frac{8}{11}\right)\rightarrow 
\left(\frac{19}{15},\frac{11}{15}\right)\rightarrow\cdots;
\nonumber\\
&&(h,\nu)=\left(\frac{7}{5},\frac{3}{5}\right)\rightarrow 
\left(\frac{4}{3},\frac{2}{3}\right)
\rightarrow \left(\frac{9}{7},\frac{5}{7}\right)\rightarrow
\cdots;\nonumber
\end{eqnarray} 

\noindent and we can verify that the other 
sequences follow according 
$h$ increases or decreases within the interval 
$1$$\;$$ < $$\;$$h$$\;$$ <$$\;$$ 2$, respecting $\nu$ 
entries in each class. The transitions allowed are 
those generated by the condition $\mid p_{2}q_{1}
-p_{1}q_{2}\mid=1$, 
with $h_{1}=\frac{p_{1}}{q_{1}}$ and $h_{2}=
\frac{p_{2}}{q_{2}}$. 
Thus our formulation satisfies in the same 
way the constraints given for the filling factors $\nu$\cite{R21}.

We have that the transition between two Hall plateaus 
is a quantum phase 
transition generated by quantum fluctuations when 
the external magnetic 
field is varied, thus we have a {\it correlation length}
 ( remember $L\sim R^h$ ) given by

\begin{equation}
\zeta\sim\frac{1}{{\mid l-l_{c}\mid}^{h}},
\end{equation}

\noindent where $l$ is the {\it magnetic length} 
which depends on 
magnetic field and temperature\footnote{The idea of the Hausdorff 
dimension as a critical exponent was discussed 
earlier in another context\cite{R22}.}. We take $h=2$, because 
we consider the {\it Hall 
fluid as a bosonic fluid}\footnote{ Observe that 
the experimental values of the 
filling factors most remarkable are just those into 
the classes $h$ between the {\it bosonic line} , $h=2$ 
and the {\it self-dual line}, $h=\frac{3}{2}$.}in according to diverse
field-theoretical models 
in the literature for FQHE\cite{R19} and better, 
experimentally the exponent of $\Delta{l}=l-l_{c}$ 
is equal to $2.02$. Also 
the dimensionless parameter $u=\frac{n\;e\;
\Delta{h}}{T^{\mu}}$ ( 
$\Delta{h}=h_{1}-h_{2}$,$\;$ $n$ is the density 
of particles with charge $e$ ) 
can be used to describe the crossover between two Hall plateaus
, with the 
{\it critical exponent} $\mu=\frac{1}{h}=0.5$ 
( assumed universal ), in according 
to the experimental value $0.45\pm 0.05$\cite{R23}. On the other hand, 
an analysis by Dolan, consider 
the critical exponent $h=2$ convenient for 
considerations about analiticity of the $\beta$-function, so the 
renormalization group parameter $u$ in this case would be 
inversely proportional to the correlation length. Thus, we can 
theoretically justify the experimental 
regularity of the transverse component of the 
conductivity at the critical point\cite{R21}.

The possible differences between the theoretical 
and experimental values for the critical exponents 
can be related with the disorder phenomenon 
and the electron-electron interactions, 
so we can consider some corrections for them. This way, we 
define $\beta-$function which describes just the 
crossover between two Hall plateaus as a complex 
analytic function of the complex conductivity

\begin{equation}
\beta(\sigma)=\frac{d\;\sigma}{d\;v},
\end{equation}

\noindent where $\sigma$ is the complex 
conductivity and $v$ is a real analytic monotonic 
function of $u$.

Following Dolan\cite{R21}, we can define

\begin{equation}
\sigma(\Delta{h})=\frac{p_{2}q_{2}\left\{K^{\prime}(w)\right\}^2+
p_{1}q_{1}\left\{K(w)\right\}^2+\imath K^{\prime}(w)K(w)}
{q_{1}^2\left\{K(w)\right\}^2+q_{2}^2\left\{K^{\prime}(w)\right\}^2},
\end{equation}

\noindent where $K^{\prime}(w)$ and $K(w)$ are complete 
elliptic integrals of the second kind, with 

\begin{equation}
w^2=\frac{1}{2}\left\{1+sign(\Delta{h})\sqrt{1-e^{-
\left(\frac{A\Delta{h}}
{\eta(\Delta{h})T^{\mu}}\right)^2}}\right\},
\end{equation}

\noindent with the complementary modulus $w^{\prime}$ defined as
 ${w^{\prime}}^2=1-w^2$, 
$K^{\prime}(w)=K(w^{\prime})$ and 
$\mu=\frac{1}{h}=0.5$ ( Hall fluid is bosonic ), 
$A$ is a positive real constant 
( which can depend on electron and impurity density, 
or other parameters )
 and the linear function $
\eta(\Delta{h})=\alpha\left\{(q_{1}-q_{2})\Delta{h}+
\alpha\right\}$, 
$\alpha=p_{2}-p_{1}-(q_{2}-q_{1})h_{c}$ and $h_{c}=h_{2}$ 
is the class 
of the filling factor $\nu_{c}=\nu_{2}$ at the critical 
value of the magnetic field.

Now, in another way, we give more arguments in favour 
of our approach. 
We have that for excitations above the 
Laughlin ground state, the exchange of two 
quasiholes\cite{R8} with coordinates $z_{\alpha}$ and 
$z_{\beta}$ produces the condition on the phase

\begin{equation}
\label{e.67}
\exp\left\{\imath\pi \nu_{1}\right\}=
\exp\left\{\imath\pi\frac{1}{m}\right\},
\end{equation}
\noindent with $\nu_{1}$$=$$\frac{1}{m}+2p_{1}$; 
and for a second generation 
of quasihole excitations, the effective wavefunction 
carries the factor $\left(z_{\alpha}-z_{\beta}
\right)^{\nu_{2}}$, 
with $\nu_{2}$$=$$\frac{1}{\nu_{1}}+2p_{2}$; 
$m=3,5,7,\cdots$ 
and $p_{1}$,$\;$$p_{2} $ are positive integers. 
We observe that these conditions 
over the filling factor $\nu$ confirm our 
classification of the collective excitations in terms 
of $h$. Another 
interesting point is that in each class, we have more 
filling factors 
which those generates by eq.(\ref{e.67}), i.e. 
our classification 
covers a more complete spectrum of states. Now, 
we can see that the duality between equivalence classes 
also means duality between quasiholes and 
quasiparticles. There are 
internal and external dualities. This means 
that for charges 
${\cal{Q}}=\pm \nu e $, we have an internal 
duality in each 
equivalence class $h$ or $\tilde{h}$. External 
duality stands 
for dual classes, 
 $h$ and $\tilde{h}$. Therefore, $h$ {\it tells us about the 
nature of the anyonic excitations}.

On one hand, we note that the anyonic exchanges of 
dual get 
a phase difference, modulo constant,

\begin{equation}
\left|\nu-\tilde{\nu}\right|=\left|\Delta\nu\right|=\left|h-
\tilde{h}\right|=const,
\end{equation}

\noindent suggesting an invariance, conformal symmetry. 
We also have for 
the elements of $h$ and $\tilde{h}$, the following relations

\begin{eqnarray}
\label{e.19}
&&\frac{\nu_{i+1}-\nu_{i}}{\tilde{\nu}_{j+1}
-\tilde{\nu}_{j}}=1;\\
&&\frac{\nu_{j+1}-\nu_{j}}{\tilde{\nu}_{i+1}
-\tilde{\nu}_{i}}=1,\nonumber
\end{eqnarray}

\noindent with $i=1,3,5,etc.$ and $j=2,4,6,etc.$; plus the pairs 
$\left(i,j\right)=\left(1,2\right),\left(3,4\right),
etc.$ satisfy the 
expressions in eq.(\ref{e.19}).

\section{Thermodynamics for fractal statistics}

The thermodynamics for 
fractal statistics is now considered and a free gas of anyons 
are termed fractons $(h,\nu)$. In particular, at low 
temperature and in the low density, for a gas with a 
constant density of states in energy, equations of state for 
such systems are obtained exactly. In particular, for particles 
in the self-dual class 
$\left\{\frac{1}{2},\frac{3}{2},\frac{5}{2},
 \frac{7}{2},\cdots\right\}_
{h={\tilde{h}}=\frac{3}{2}}$, 
we obtain the distribution

\begin{equation}
n=\frac{1}{\sqrt{\frac{1}{4}+\xi^2}},
\end{equation}

\noindent and we  {\it observe} that this expression was just obtained 
in the literature for the statistics parameter $g=\frac{1}{2}$\cite{R13}, 
but {\it according to our approach} it is valid for all values of $\nu$ 
within the class $h=\frac{3}{2}$. It is noteworthy that for some 
of these special values of $\nu$ 
( with even denominators ) FQHE were observed\cite{R24}.  

The thermodynamic properties for a given class $h$ 
can be now obtained. In two dimensions, the particle number in a 
volume element ${d^2}p\;dV$ in phase space is given by

\begin{eqnarray}
dN={\cal{G}}\frac{d^2p\;dV}{(2\pi {\hbar})^2}
\frac{1}{{\cal{Y}}(\xi)-h},
\end{eqnarray}

\noindent with ${\cal{G}}=\nu+1$ and $\nu=2s$. Integrating over
 $V(area)$, we obtain

\begin{eqnarray}
dN_{p}=\frac{{\cal{G}}\;V}{2\pi {\hbar}^2}
\frac{p\;dp}{{\cal{Y}}(\xi)-h},
\end{eqnarray}

and for the dispersion $\epsilon(p)=\frac{p^2}{2m}$

\begin{eqnarray}
dN_{\epsilon}=\frac{m\;{\cal{G}}\;V}{2\pi{\hbar}^2}
\frac{d\epsilon}{{\cal{Y}}[\xi(\epsilon)]-h}.
\end{eqnarray}

The total energy is given by

\begin{eqnarray}
E&=&\int_{0}^{\infty}\epsilon\; dN_{\epsilon}\nonumber\\
&=&\frac{m\;{\cal{G}}\;V}{2\pi{\hbar}^2}\int_{0}^{\infty}
\frac{\epsilon\; d\epsilon}{{\cal{Y}}[\xi(\epsilon)]-h}
\end{eqnarray}

\noindent and  

\begin{equation}
PV=E=\frac{m\;{\cal{G}}\;V}{2\pi{\hbar}^2}\int_{0}^{\infty}
\frac{\epsilon\; d\epsilon}{{\cal{Y}}[\xi(\epsilon)]-h}.
\end{equation}

\noindent For the thermodynamic potential we obtain

\begin{equation}
\Omega=-PV=-\frac{m\;{\cal{G}}\;V}{2\pi{\hbar}^2}\int_{0}^{\infty}
\frac{\epsilon\;d\epsilon}{{\cal{Y}}[\xi(\epsilon)]-h}
\end{equation}

\noindent and for the class $h=\frac{3}{2}$ and 
$\nu=\frac{1}{2}$, the pressure is given by

\begin{equation}
P=\frac{3}{2}\frac{m}{2\pi{\hbar}^2}\int_{0}^{\infty}
\frac{\epsilon\;d\epsilon}
{\sqrt{\frac{1}{4}+\exp{2(\epsilon-\mu)/KT}}};
\end{equation}

\noindent defining the variable $z=\frac{\epsilon}{KT}$, 
we written down 

\begin{equation}
\frac{P}{KT}=\frac{3}{2}\frac{m\;K\;T}{2\pi{\hbar}^2}\int_{0}^{\infty}
\frac{2\;z\;dz}
{\sqrt{1+4\exp{2(z-\mu/KT)}}},
\end{equation}

\noindent which combined with the particle density

\begin{equation}
\frac{N}{V}=\frac{3}{2}\frac{m\;K\;T}{2\pi{\hbar}^2}\int_{0}^{\infty}
\frac{2\;dz}
{\sqrt{1+4\exp{2(z-\mu/KT)}}},
\end{equation}

\noindent determines the relation between $P,V$ and $T$, the
 equation of state for a gas of free particles termed {\it fracton} $(h,\nu)
 =\left(\frac{3}{2},\frac{1}{2}\right)$.
 
 From the eq.(\ref{e.46}) we see that the function ${\cal{Y}}(\xi)$ get 
 its degree from $h$ denominator and so for calculate the roots of 
 ${\cal{Y}}$ in terms of the parameter $\xi$ can be a 
 formidable mathematical task. Otherwise, we can solve ${\cal{Y}}(\xi)$ 
 numerically. For example, the classes $h=\frac{4}{3},\frac{5}{3}$ have 
 a third degree algebraic equation
 
\begin{equation}
\label{e.16}
{{\cal{Y}}}^3+a_{1}{{\cal{Y}}}^2+a_{2}{\cal{Y}}+a_{3}=0,
\end{equation}

\noindent which has real solution ( Cardano formula )

\begin{equation}
\label{e.17}
{\cal{Y}}(\xi)=s+t-\frac{a_{1}}{3},
\end{equation}

\noindent where

\begin{eqnarray}
\label{e.18}
&&s=\left\{{\sqrt{r+{\sqrt{q^3+r^2}}}}\right\}^{3},\nonumber\\
&&t=\left\{{\sqrt{r-{\sqrt{q^3+r^2}}}}\right\}^{3}\nonumber
\end{eqnarray}

\noindent and

\begin{eqnarray}
&&q=\frac{3a_{2}-{a_{1}}^2}{9},\nonumber\\
&&r=\frac{9a_{1}a_{2}-27a_{3}-2{a_{1}}^3}{54}.\nonumber
\end{eqnarray}

\noindent For the class $h=\frac{4}{3}$ we obtain

\begin{equation}
{{\cal{Y}}}^3-5{{\cal{Y}}}^2+8{\cal{Y}}-(4+\xi^{3})=0,
\end{equation}

\noindent and for $h=\frac{5}{3}$

\begin{equation}
{{\cal{Y}}}^3-4{{\cal{Y}}}^2+5{\cal{Y}}-(2+\xi^{3})=0.
\end{equation}

\noindent The average ocuppation number is given by

\begin{equation}
n=\frac{1}{s+t-\frac{1}{3}a_{1}-h}
\end{equation}

\noindent and the thermodynamic properties for the 
{\it fractons} $(h,\nu)=\left(\frac{4}{3},\frac{2}{3}\right);
\left(\frac{5}{3},\frac{1}{3}
\right)$, can be considered. However, 
we can follow another way. The 
distribution function can be written 
in terms of the single-state grand 
partition function ${\Theta}_{j}$, as

\begin{equation}
n_{j}=\xi_{j}\frac{\partial\ln\Theta_{j}}{\partial{{\xi}_{j}}},
\end{equation}

\noindent where $\Theta_{j}=\frac{{\cal{Y}}_{j}-2}{{\cal{Y}}_{j}-1}$. 
Thus, we can expand $\Theta_{j}$ in powers of $\xi_{j}$,

\begin{eqnarray}
\Theta_{j}&=&\frac{{\cal{Y}}_{j}-2}{{\cal{Y}}_{j}-1}\nonumber\\
&=&\sum_{l=0}^{\infty}V_{l}\;\xi_{j}^{l}
\end{eqnarray} 

\noindent and

\begin{eqnarray}
n_{j}=\sum_{l=1}^{\infty}U_{l}\;\xi_{j}^{l},
\end{eqnarray}

\noindent where

\begin{eqnarray}
V_{l}=\prod_{k=2}^l\left\{1+(h-2)\frac{l}{k}\right\};\;
U_{l}=\prod_{k=1}^{l-1}\left\{1+(h-2)\frac{l}{k}\right\}.\nonumber
\end{eqnarray}

\noindent These coefficients are obtained 
in the same way as discussed in\cite{R25}. They do not depend on 
$G$ states number and are calculated from the 
combinatorial formula  
 
\begin{equation}
\label{e.23}
w=\frac{G\left[G+(h-1)N-1\right]!}{N!
\left[G+(h-2)N\right]!},
\end{equation}

\noindent taking into account the grand partition function
 
\begin{eqnarray}
\label{e.1}
{\cal{Z}}(G,\xi)&=&\sum_{N=0}^{\infty}w(G,N)\;\xi^N\\
&=&\left(\sum_{l}V_{l}\;\xi^l\right)^G\nonumber,
\end{eqnarray}

\noindent as $G$-th power of the 
single-state partition function $\Theta$. 

The eq.(\ref{e.23}) 
gives the same statistical mechanics as 
that introduced earlier in eq.(\ref{e.15}). 
Now, we can consider at low temperature Sommerfeld 
expansions\cite{R26}. This way, we handle 
integrals of the type

\begin{eqnarray}
{\cal{J}}[f]=KT\int_{1}^{\infty}\frac{d{\cal{Y}}}{\left({\cal{Y}}
-1\right)\left({\cal{Y}}-2\right)}
\;f\left\{\mu+KT
\left[h\ln\left\{\frac{{\cal{Y}}-1}{{\cal{Y}}-2}\right\}
-\ln\left\{\frac{{\cal{Y}}-1}{\left({\cal{Y}}-2\right)^2}
\right\}\right]\right\},
\end{eqnarray}

\noindent and after expansion we have

\begin{eqnarray}
{\cal{J}}[f]=h^{-1}\int_{0}^{\mu}\;f(\epsilon)
\;d\epsilon+\sum_{l=0}^{\infty}
\frac{(KT)^{l+1}}{l!}C_{l}(h)f^{(l)}(\mu),
\end{eqnarray}

\begin{eqnarray}
C_{l}(h)=\sum_{k=0}^{l-1}C_{l,k}\;h^k,
\end{eqnarray}

\noindent with

\begin{eqnarray}
C_{l,k}=(-)^{l-k}
\left(\begin{array}{c}
l\\k
\end{array}\right)
\int_{1}^{\infty}\frac{d{\cal{Y}}}{\left({\cal{Y}}-1\right)
\left({\cal{Y}}-2\right)}
\ln^{l-k}\left\{\frac{{\cal{Y}}-1}{{\cal{Y}}-2}\right\}\ln^{k}
\left\{\frac{{\cal{Y}}-1}
{({\cal{Y}}-2)^2}\right\}.
\end{eqnarray}

We can obtain now some thermodynamic quantities. 
Let us consider, 
the dispersion as $\epsilon(p)=ap^\sigma$, and 
in D-dimension we get for a few 
terms of $C_{l}$, the expressions

\begin{eqnarray}
\frac{\cal{E}}{\gamma\;V}=\frac{\mu^{\frac{D}{\sigma}+1}}
{h\left(\frac{D}{\sigma}+1\right)}+
KTC_{0}(h)\mu^{\frac{D}{\sigma}}+\frac{1}
{2}(KT)^2C_{1}(h)
\left(\frac{D}{\sigma}\right)\mu^{\frac{D}{\sigma}-1}+\cdots,
\end{eqnarray}

\begin{eqnarray}
\frac{{\cal{N}}}{\gamma\;V}=\frac{\mu^{\frac{D}{\sigma}}}
{h\left(\frac{D}{\sigma}\right)}+
KTC_{0}(h)\mu^{\frac{D}{\sigma}-1}+\frac{1}
{2}(KT)^2C_{1}(h)
\left(\frac{D}{\sigma}-1\right)\mu^{\frac{D}{\sigma}-2}+\cdots,
\end{eqnarray}

\noindent for the energy and the 
particle number, where $\gamma=\frac{m\;{\cal{G}}}{4\pi\hbar^2}$,
 for $D=\sigma=2$. The coefficients 
 $C_{0}(h)$,$C_{1}(h)$ are given by

\begin{eqnarray}
C_{0}(h)&=&C_{0,0}=\int_{1}^{\infty}\frac{d{\cal{Y}}}
{({\cal{Y}}-1)({\cal{Y}}-2)};\\
C_{1}(h)&=&C_{1,0}=-\int_{1}^{\infty}\frac{d{\cal{Y}}}
{({\cal{Y}}-1)({\cal{Y}}-2)}
\ln\left\{\frac{{\cal{Y}}-1}{{\cal{Y}}-2}\right\}.\nonumber
\end{eqnarray}

\noindent We can show that to second order $T$

\begin{eqnarray}
&&{\cal{E}}={\cal{E}}_{0}\left\{1+h\;\frac{KT}{\mu_{0}}C_{0}(h)+
2h\;\frac{(KT)^2}{\mu_{0}^2}C_{1}(h)\left(\frac{D}{\sigma}\right)
-h\;C_{0}^2(h)\frac{(KT)^2}{\mu_{0}^2}
\left(\frac{D}{\sigma}\right)
\left(\frac{D}{\sigma}+1\right)\right\},\\
&&\mu=\mu_{0}\left\{1-h\;\frac{KT}{\mu_{0}}
\left(\frac{D}{\sigma}\right)C_{0}(h)
-h\;\frac{(KT)^2}{\mu_{0}^2}C_{1}(h)
\left(\frac{D}{\sigma}\right)\left(\frac{D}
{\sigma}-1\right)\right\},
\end{eqnarray}

\noindent for the energy and chemical potential, 
respectively; and 
to order $T$

\begin{eqnarray}
&&{\cal{C}}(h)=h\;{\cal{E}}_{0}\frac{K}{\mu_{0}}C_{0}(h)
+2h\;{\cal{E}}_{0}\frac{K^2T}{\mu_{0}^2}C_{1}(h)\left(
\frac{D}{\sigma}\right)-h\;
{\cal{E}}_{0}\frac{K^2T}{\mu_{0}^2}C_{0}^2(h)
\left(\frac{D}{\sigma}\right)
\left(\frac{D}{\sigma}+1\right),
\end{eqnarray}

\noindent for the specific heat, where 
$\mu_{0}$ and ${\cal{E}}_{0}$ are the 
zero-temperature chemical potential and energy.

For a dilute gas or high-temperature regime, we can 
investigate the thermodynamic properties of a 
free gas using virial expansion. The 
equation of state relates $P$, $T$ and 
$\rho$ ( pressure, temperature 
and density respectively ) as follows\cite{R27}

\begin{equation}
P=\rho\;K\;T\left\{1+a_{2}\rho+a_{2}\rho^2+\cdots\right\},
\end{equation}

\noindent where $a_{2}$,$a_{3}$, etc are 
the virial cofficients. The 
grand potential is the sum over the single-particle states

\begin{equation}
\Omega=-KT\sum_{j}\ln\Theta_{j},
\end{equation}

\noindent and as $\Omega=-PV$, we have

\begin{equation}
\frac{P}{KT}=\sum_{j}\frac{1}{V}\ln\Theta_{j},
\end{equation}

\noindent and expanding in terms of $\xi_{j}$ 
after summation over $j$, we get the cluster expansion

\begin{equation}
P=KT\sum_{l=1}^{\infty}b_{l}\;z^{l},
\end{equation}

\noindent and for density

\begin{equation}
\rho=\sum_{l=1}^{\infty}l\;b_{l}\;z^{l},
\end{equation}

\noindent where $b_{l}=\prod_{k=1}^{l-1}
\left\{1+(h-2)\frac{l}{k}\right\}
\frac{{\cal{Z}}_{1}(\frac{T}{l})}{l}$,  
$z=\exp\frac{\mu}{KT}$ is the fugacity and 
${\cal{Z}}_{1}(T)=V\gamma(KT)^{\frac{D}{\sigma}}$ 
is the one-particle partition function $\sum_{j}
\exp\left\{-\frac{\epsilon_{j}}{KT}\right\}$. This 
way, we can show that

\begin{equation}
a_{2}=2^{-\frac{D}{\sigma}}\left(h-\frac{3}{2}\right)\frac{V}
{{\cal{Z}}_{1}(T)},
\end{equation}

\begin{equation}
a_{3}=\left\{\left(5-3h\right)\left(3h-4\right)
3^{-(\frac{D}{\sigma}+1)}+
2^{-2(\frac{D}{\sigma})}\left(2h-3\right)^2 
\right\}\left[\frac{V}
{{\cal{Z}}_{1}(T)}\right]^2,
\end{equation}

\noindent with $a_{l}={\tilde{a}}_{l}V^{l-1}$,$\;$ 
${\tilde{b}}_{l}=\frac{b_{l}}{b_{1}^l}$,$\;$ 
${\tilde{a}}_{2}=-{\tilde{b}}_{2}$,$\;$ 
${\tilde{a}}_{3}=-2{\tilde{b}}_{3}+4{\tilde{b}}_{2}^2$, etc.

For a gas with a constant density of states in energy, 
with $D=\sigma=2$, we have

\begin{equation}
\frac{{\cal{Y}}(0)-1}{{\cal{Y}}(0)-2}=e^{-\frac{\mu}{hKT}}=
e^{-\frac{\rho}{\gamma\;KT}}
\end{equation}

\noindent and from the eq.(\ref{e.46}) we obtain

\begin{equation}
\mu(\rho,T)=\frac{(h-1)\rho}{\gamma}+KT\ln\left\{e^{-\frac{\rho}
{\gamma\;KT}}-1\right\}
\end{equation}

\noindent and in the low density, the pressure is a finite expression

\begin{equation}
P=\left(h+1\right)\frac{\rho^2}
{2\gamma}
\end{equation}

\noindent and at low temperature, we obtain

\begin{equation}
P=\frac{h\rho^2}{2\gamma}+\gamma(KT)^2C_{1}(h).
\end{equation}

\noindent As a consequence, the specific heat depends on $h$ and $\nu$.

\section{Farey sequences and Hausdorff dimension}

From our considerations about Fractional 
Quantum Hall Effect we can show now a connection 
between the fractal parameter $h$ and the Farey series 
for rational numbers\cite{R28}. 
Thus, we have the following {\bf theorem}:

{\it The elements of the Farey series belong to 
distinct equivalence classes labeled by a fractal parameter $h$
defined into the interval 
$1$$\;$$ < $$\;$$h$$\;$$ <$$\;$$ 2$, and these classes satisfy 
the same properties observed for those fractions. Also, 
for each value of $h$ there exists an algebraic equation }.

As we have pointed out above the fractal parameter 
$h$ is related to $\nu$ ( an 
irreducible number $\frac{p}{q}$, 
with $p$ and $q$ integers, see eq.(\ref{e34}) ) and 
we can extract, for example, the classes

\begin{eqnarray}
&&\left\{\frac{1}{3},\frac{5}{3},\frac{7}{3},
 \frac{11}{3},\cdots\right\}_
{h=\frac{5}{3}},\;\;\;\;\;\;\;\;\;\;\;\;\;\;\left\{\frac{5}{14},
\frac{23}{14},\frac{33}{14},\frac{51}{14},\cdots\right\}_
{h=\frac{23}{14}},\nonumber\\
&&\left\{\frac{4}{11},\frac{18}{11},\frac{26}{11},
 \frac{40}{11},\cdots\right\}_
{h=\frac{18}{11}},\;\;\;\;\;\;\;\;\left\{\frac{7}{19},
\frac{31}{19},\frac{45}{19},\frac{69}{19},\cdots\right\}_
{h=\frac{31}{19}},\nonumber\\
&&\left\{\frac{10}{27},\frac{44}{27},\frac{64}{27},
 \frac{98}{27},\cdots\right\}_
{h=\frac{44}{27}},\;\;\;\;\;\;\;\;\left\{\frac{3}{8},
\frac{13}{8},\frac{19}{8},\frac{29}{8},\cdots\right\}_
{h=\frac{13}{8}},\\
&&\left\{\frac{3}{7},\frac{11}{7},\frac{17}{7},
\frac{25}{7},\cdots\right\}_
{h=\frac{11}{7}},\;\;\;\;\;\;\;\;\;\;\left\{\frac{4}{9},
\frac{14}{9},\frac{22}{9},\frac{32}{9},
\cdots\right\}_{h=\frac{14}{9}},\nonumber\\
&&\left\{\frac{5}{11},\frac{17}{11},\frac{27}{11},
\frac{39}{11},\cdots\right\}_
{h=\frac{17}{11}},\;\;\;\;\;\;\;\;\left\{\frac{6}{13},
\frac{20}{13},\frac{32}{13},
\frac{46}{13},\cdots\right\}_{h=\frac{20}{13}},\nonumber\\
&&\left\{\frac{2}{5},\frac{8}{5},\frac{12}{5},
\frac{18}{5},\cdots\right\}_
{h=\frac{8}{5}}\nonumber
\end{eqnarray}

\noindent and for that we can consider the 
serie $(h,\nu)$

\begin{eqnarray}
&&\left(\frac{5}{3},\frac{1}{3}\right)\rightarrow 
\left(\frac{18}{11},\frac{4}{11}
\right)\rightarrow
\left(\frac{13}{8},\frac{3}{8}\right)\rightarrow
 \left(\frac{8}{5},\frac{2}{5}\right)\rightarrow\\
&&\left(\frac{11}{7},\frac{3}{7}\right)\rightarrow 
\left(\frac{14}{9},\frac{4}{9}\right)
\rightarrow \left(\frac{17}{11},\frac{5}{11}\right)
\rightarrow 
\left(\frac{20}{13},\frac{6}{13}\right) \rightarrow 
\cdots\nonumber
\end{eqnarray} 

\noindent The classes $h$ satisfy all properties of the 
Farey series: 

P1. If $h_{1}=\frac{p_{1}}{q_{1}}$ and 
$h_{2}=\frac{p_{2}}{q_{2}}$ are two consecutive fractions 
$\frac{p_{1}}{q_{1}}$$ >$$ \frac{p_{2}}{q_{2}}$, then 
$|p_{2}q_{1}-q_{2}p_{1}|=1$.

P2. If $\frac{p_{1}}{q_{1}}$, $\frac{p_{2}}{q_{2}}$,
$\frac{p_{3}}{q_{3}}$ are three consecutive fractions 
$\frac{p_{1}}{q_{1}}$$ >$$ \frac{p_{2}}{q_{2}} 
$$>$$ \frac{p_{3}}{q_{3}}$, then 
$\frac{p_{2}}{q_{2}}=\frac{p_{1}+p_{3}}{q_{1}+q_{3}}$.

P3. If $\frac{p_{1}}{q_{1}}$ and $\frac{p_{2}}{q_{2}}$ are 
consecutive fractions in the same sequence, then among 
all fractions\\
 between the two, 
$\frac{p_{1}+p_{2}}{q_{1}+q_{2}}$
 is the unique reduced
fraction with the smallest denominator.
 
For more details about Farey series see\cite{R28}. 
All these properties can be verified for the 
classes considered above as an example. Another 
one is

\begin{eqnarray}
(h,\nu)&=&\left(\frac{11}{6},\frac{1}{6}\right)\rightarrow 
\left(\frac{9}{5},\frac{1}{5}
\right)\rightarrow
\left(\frac{7}{4},\frac{1}{4}\right)\rightarrow
 \left(\frac{5}{3},\frac{1}{3}\right)\rightarrow\\
&&\left(\frac{8}{5},\frac{2}{5}\right)\rightarrow 
\left(\frac{3}{2},\frac{1}{2}\right)
\rightarrow \left(\frac{7}{5},\frac{3}{5}\right)
\rightarrow 
\left(\frac{4}{3},\frac{2}{3}\right) \rightarrow\nonumber\\ 
&&\left(\frac{5}{4},\frac{3}{4}\right) \rightarrow 
\left(\frac{6}{5},\frac{4}{5}\right) \rightarrow 
\left(\frac{7}{6},\frac{5}{6}\right) \rightarrow 
\cdots,\nonumber
\end{eqnarray}

\noindent where the $\nu$ sequence is 
the Farey series of order $6$. Thus, we observe 
that because of the {\it fractal spectrum} in eq.(\ref{e34}), 
we can write down any Farey series of rational numbers. 
Therefore, in this way, we established a beautiful connection 
between number theory and physics. We have shown that 
the Farey series can be arranged into equivalence 
classes labeled by a fractal parameter $h$ 
which looks like a Hausdorff 
dimension. For each value of $h$ we have an algebraic 
equation derived of the eq.(\ref{e.46}). Then, there exists 
a relation between algebraic equation and Farey series. The 
connection between a geometric parameter related to the paths 
of particles  and rational numbers is for itself 
an interesting result.

\section{summary and discussion}

In summary, we have obtained a path integral representation for the 
propagator of free anyons eq.(\ref{a12}) and its 
representation in momentum space eq.(\ref{a7}) taking into 
account a continuous family of Lagrangians eq.(\ref{a14}). 
These ones 
were obtained by a group-theoretical approach with minimal 
extension that preserves all canonical structure 
of the space-time and spin algebra\cite{R3}. To 
obtain the propagator, 
a convenient gauge ${\dot{\sigma}}$=$0$,$\;$${\dot{e}}$=$0$ 
was considered. 
We have furthermore obtained distribution 
functions eq.(\ref{e.45}) for 
anyonic excitations in terms of the Hausdorff dimension 
$h$, which classifies the anyonic excitations into 
equivalence classes ( or universal classes of particles, 
because we have 
{\it fractons}$\;$$(h,\nu)$) and reduces to 
fermionic and bosonic distributions, when $h=1$ and 
$h=2$, respectively. This constitutes a new approach 
for such systems of fractional spin particles. 
This way, we have extended eq.(\ref{e.4}) results 
of the literature\cite{R13} 
for the complete spectrum of 
statistics $\nu$. 

A connection with the FQHE, 
considering the filling factors into equivalence classes 
labeled by $h$ was also
considered and an estimate for occurrence 
of FQHE made. A relation 
between equivalence classes $h$ and the modular group for 
the quantum phase transitions ( meaning that $h$ classifies 
the universality class of these transitions ) 
of the FQHE was also noted. 
A $\beta-$function is written down\cite{R21} which 
embodies the classes $h$ and the 
concept of duality between equivalence classes confirms that 
all this works. 

We have also considered a free gas 
of fractons $(h,\nu)$ and an exact equation of state was obtained 
at low-temperature and low-density limits. In particular, 
for a gas in the low-density regime the equation of state shows us 
only interaction between pairs of group of two clusters 
of fractons $(h,\nu)$. A connection between Farey 
sequences and Hausdorff dimension have been proven. We have shown 
that the idea of supersymmetry appears naturally in 
solid state physics, throughout our 
approach to FQHE\cite{R20}. This was possible with the introduction of 
the concept of duality between equivalence classes defined by
 ${\tilde{h}}-1=2-h $ or 
$ h-1=2-{\tilde{h}} $. These coefficients take part 
in eq.(\ref{e.46}) as exponents. It is worth noting that the fractal 
parameter $h$ can be any number within the interval 
of definition and so we have other possibilities for 
the spin values to the particles in two space dimensions. 
Another important theoretical point is that we considered 
{\it ab initio} the spin-statistics relation $\nu=2s$. 
Observe that our results are supported by the symmetries 
which pervade all the formulation propose by us.

Now, a note of caution: The filling factor is defined by 
$f=N\frac{\phi_{0}}{\phi}$ ( where $N$ is the number of electrons, 
$\phi_{0}$ is the quantum unit of magnetic flux and 
$\phi$ is the external magnetic flux through the sample material ) and 
the spin-statistics relation is given by 
$\nu=2s=2\frac{\phi\prime}{\phi_{0}}$ ( where 
$\phi\prime$ is the magnetic flux 
associated to the charge-flux system ). According to our approach 
there is a correspondence between $f$ and $\nu$, 
numerically $f=\nu$. Thus, the quantum Hall states 
with $f$ taking odd, even integer and rational values 
would be described, 
respectively by fermions ($h=1$), bosons ($h=2$) and 
fractons ($1$$\;$$ < $$\;$$h$$\;$$ <$$\;$$ 2$)\footnote{Keep 
in mind the dual character of our approach, $\tilde{h}=3-h $.}. This way, 
we have a more general setting for particles or 
quasiparticles termed fractons $(h,\nu)$, with $1\leq h\leq 2$. 
As a result, we get some information about the microscopic nature of 
both phenomena: The integer and fractional quantum Hall effect. Another 
interesting conceptual possibility is the {\it pairing of fractons $(h,\nu)$}, 
resulting in fermions, bosons or fractons. A target of these ideas can be, for 
example, the high-$T_{c}$ superconductivity\cite{R30}.

 Finally, 
we would like {\it to emphasize} that in our 
approach to fractional spin particles we do not have 
any statistical interaction matrix which describes 
correlation effects among particles according to the 
generalized Pauli exclusion principle by Haldane\cite{R11}. However, 
as we have seen the fractal parameter $h$ gives us distributions which 
coincide with Haldane analysis, but contrary to him and 
previous authors\cite{R12,R13,R14,R29} our interpretation is that 
such distributions are valid for all particles in 
a specific class $h$. In this way, we 
generalize the Fermi-Dirac and Bose-Einstein statistics, 
taking into account {\it a unique} expression eq.(\ref{e.45}) 
for each universal class $h$ of particles. 
This is {\bf the difference}. From a mathematical viewpoint 
we have also discovered all a class of {\it fractal functions}, i.e. 
just our fractal statistics ( see Rocco and West in\cite{R6} ). This one
captures the observation that the path of a quantum-mechanical particle is 
continuous and nondifferentiable, i.e. a fractal curve. Hence, our 
formulation can be understood as a {\it quantum-geometrical} 
description of the statiscal laws of Nature.  
It is noteworthy that this result is a physical 
realization of mathematical ideas about fractal geometry. 
An extension of this 
abelian fractal statistics to nonabelian fractal statistics, 
the case for the higher dimensional irreducible unitary 
representations of the braid 
group is a matter of future work. Also the study of some 
connection between effective conformal field 
theories for quasiparticles and these statistics 
deserves attention\cite{R31,R32}. 
Much of the results of this paper appeared 
in a series of unpublished works by the author.

\end{document}